\documentclass[doublecol]{epl2} 
\usepackage{psfrag}
\usepackage{amsmath}
\usepackage{amssymb}

\newcommand{\half}{\textstyle \frac{1}{2}}

\title{Transport signature of pseudo-Jahn-Teller dynamics in a single-molecule transistor}
\author{F. Reckermann\inst{1,2,3} \and M. Leijnse\inst{1,3} \and M. R. Wegewijs\inst{1,2,3} \and H. Schoeller\inst{1,3}}

\institute{                    
  \inst{1} Institut f\"ur Theoretische Physik A, RWTH Aachen, 52056 Aachen,  Germany\\
  \inst{2} Institut f\"ur Festk{\"o}rper-Forschung - Theorie 3, Forschungszentrum
  J{\"u}lich, 52425 J{\"u}lich,  Germany\\
  \inst{3} JARA- Fundamentals of Future Information Technology
}


\abstract{
  We calculate the electronic transport through a molecular dimer,
  in which an excess electron is delocalized over equivalent monomers, which can be locally distorted.
  In this system the Born-Oppenheimer approximation breaks down
  resulting in quantum entanglement of the mechanical and electronic motion.
  We show that pseudo Jahn-Teller (pJT) dynamics of the molecule
  gives rise to conductance peaks that indicate this violation.
  Their magnitude, sign and position sharply depend on the electro-mechanical properties of the molecule,
  which can be varied in recently developed three-terminal junctions with mechanical control.
  The predicted effect depends crucially on the degree of intramolecular delocalization of the excess electron,
  a parameter which is also of fundamental importance in physical chemistry.}

\begin{document}

\maketitle

\section{Introduction}

% NEMS
Nano-electromechanical devices (NEMS) electrically detect and control mechanical motion with great precision~\cite{Ekinci05}
and can be constructed in various nanostructures, including macromolecules such as suspended carbon nanotubes~\cite{Sazonova04,Sapmaz05}.
% GATED MOLECULES
Nowadays  even nano-meter sized molecules are within reach of experimental investigation.
Successful three-terminal transport measurements~\cite{Park99elmig} have been reported,
detecting the quantized vibrational~\cite{Park00,Pasupathy04},
spin~\cite{Osorio07b} and magnetic~\cite{Heersche06,Jo06} excitations of a single molecule.
The quantum-limited operation of NEMS is thus a starting point, rather than a goal in the single-molecule regime.
More challenging is achieving control over such devices.
Recently, electrical three-terminal devices have been demonstrated with additional mechanical control~\cite{Champagne05,Parks07}.
Here the size of the nanogap in which the molecule is embedded can be adjusted with sub-\AA ngstrom precision,
thereby changing the capacitive and resistive coupling of the molecule to the source and drain electrodes,
as well as the intrinsic molecular properties.
%
% Molecular QEMS
The interesting question arises how single-molecule quantum states involving electronic and mechanical degrees of freedom
may be detected and controlled in such transport experiments.
% 1 Adiabatic, FC effects
This has been addressed in several theoretical studies, e.g.~\cite{Flensberg03,Mitra04b,Koch04b,Wegewijs05}.
Fundamental to nearly all of these works is the adiabatic Born-Oppenheimer (BO) approximation, where one separates
the timescales of the fast electronic motion from the slow dynamics of the nuclei.
The transport is then governed by the Franck-Condon (FC) principle,
where the tunneling of an electron onto the molecule
changes the electron number $N\rightarrow N+1$,
which induces a transition between electronic states $e \rightarrow e'$
and the initial vibrational state $\chi_{e}$ is projected onto
 a vibrational state $\chi'_{e'}$ of the molecule in the final electronic state.
The amplitude for this process factorizes and is proportional to the overlap
integrals of the mechanical wave functions $\langle\chi_{e}|\chi'_{e'}\rangle$
which in general is non-zero and strongly depends on both of the vibrational states.
There are thus no selection rules in contrast to spin-related tunneling.
By proper choice and design of the vibrational properties of molecular transistors
(number of modes, adiabatic potential landscapes, etc.),
the FC-effect may thus be exploited to induce non-trivial vibrational states
and interesting non-linear transport characteristics
relevant for possible electro-mechanical sensing applications.
\\
% 2 
A novel aspect of molecule-based NEMS is they may display strong \emph{vibronic} effects (distinct from vibrational)
due to the non-trivial coupled quantum dynamics of the electronic  and nuclear degrees of freedom,
for a review see~\cite{Bersuker89}.
Here the Born-Oppenheimer separation of the time-scales
for the nuclear and electron motion breaks down.
The system is only adequately described by so-called vibronic states, in which
the quantum entanglement renders the concept of electronic and nuclear motion meaningless.
The most prominent and well studied vibronic effect is the \emph{dynamical} Jahn-Teller (JT) effect, 
which occurs in molecules where the electronic ground state is degenerate due to a high spatial symmetry of the static nuclear framework of the (non-linear) molecule.
In such systems, there always exists~\cite{Jahn37} a vibrational coordinate
along which a static distortion will lower the molecular symmetry and lift the degeneracy.
However, in a single molecule, the distortions are dynamical and
the electronic degeneracy is transformed into a vibronic degeneracy i.e. of the quantum-mechanical \emph{molecular} eigenstates.
%
% Jahn-Teller and transport
Recently, the selection rules encoded in these molecular eigenstates (related to the high symmetry)
were predicted to block electron transport through a JT active molecule~\cite{Schultz07a}.
% Issues / questions raised
% 1
An important question now is how to distinguish such vibronic blockade from 
spin-~\cite{Romeike07a}, magnetic~\cite{Romeike06b} or Franck-Condon~\cite{Braig03a,Braig04b,Koch04b} blockade effects.
% 2
Another issue is that the BO-approximation breaks down
even when electronic levels only come close in energy
(on the order of a few vibrational energy spacings).
This is referred to as the dynamical \emph{pseudo Jahn-Teller} (pJT) effect
and occurs in many molecular systems~\cite{Bersuker89}.
A generic problem where it occurs is in determining
the degree of delocalization of an excess electron in molecular mixed-valence compounds~\cite{Bersuker92},
which is fundamental to the classification by the Robin-Day (RD) scheme in physical chemistry~\cite{RobinDay67}.
The dynamical pJT effect is relevant in any system where electrons become 
delocalized over multiple similar centers, while local vibrational modes couple to the charge on the centers.
Molecular dimers, being of recent experimental interest~\cite{Chang07, Pasupathy04}, 
constitute a basic system which may exhibit this effect.
Such a system may be called ``molecular double quantum dot''.
However, due to the vibrations and the pJT effect
its transport properties dramatically deviate from semiconductor double quantum dots.
More generally, such mixed-valence effects are important in functional nanosystems 
such as metal-organic supramolecular arrays~\cite{Ruben04}
and polyoxometalates, see e.g.~\cite{Borshch06}.
\\
In this letter, we
predict transport signatures of the pJT dynamics,
of a single molecular dimer transistor,
which markedly differ from those due to the Franck-Condon effect,
allowing the breakdown of the Born-Oppenheimer principle to be identified experimentally.
We find characteristic non-linear conductance peaks with a sharp dependence of their
\emph{position, magnitude and sign} on the electro-mechanical parameters of the molecule.

\section{Model}
We consider a dimer molecule consisting of two identical monomers, labelled by $i=1,2$.
Each monomer can vibrate along its individual totally symmetric (``breathing'') mode $Q_i$ about the potential minimum at $Q_i=0$ with frequency $\omega$.
Each monomer also accomodates one electronic orbital state, $\vert i \sigma\rangle$, for an excess electron with spin projection $\sigma$
that can tunnel to the other monomer with amplitude $t$ via a mechanically stiff bridging ligand.
\begin{figure}[htbp]
  \psfrag{gap}[c][c][0.7][0.0]{gap $=2t$}
  \psfrag{wg}[b][b][0.7][0.0]{$\omega_g$}
  \psfrag{we}[b][b][0.7][0.0]{$\omega_e$}
  \psfrag{Qmin}[l][l][0.5][0.0]{$Q_{-}$}
  \centering
  \begin{tabular}{c}
    \begin{tabular}{c}
      (a)\, \includegraphics[width=7cm]{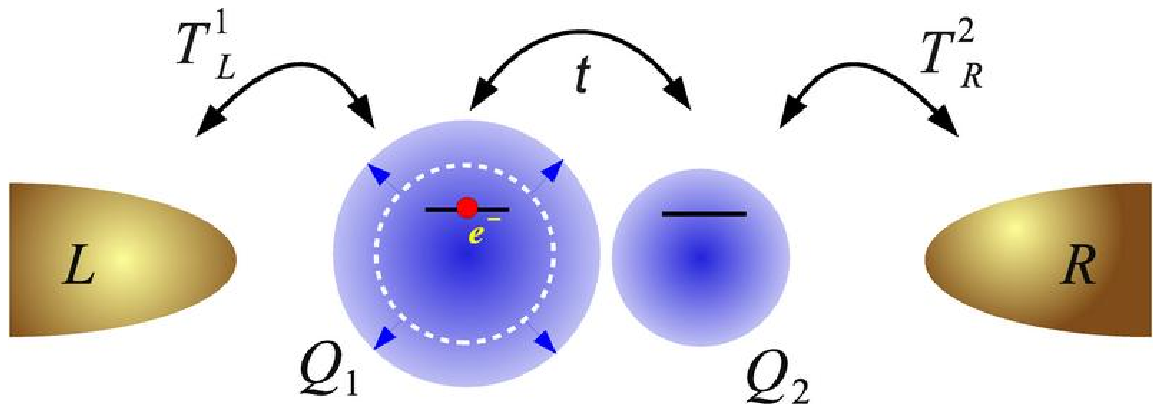}\\
      \begin{tabular}{ccc}
        (b)\, \includegraphics[width=2.1cm]{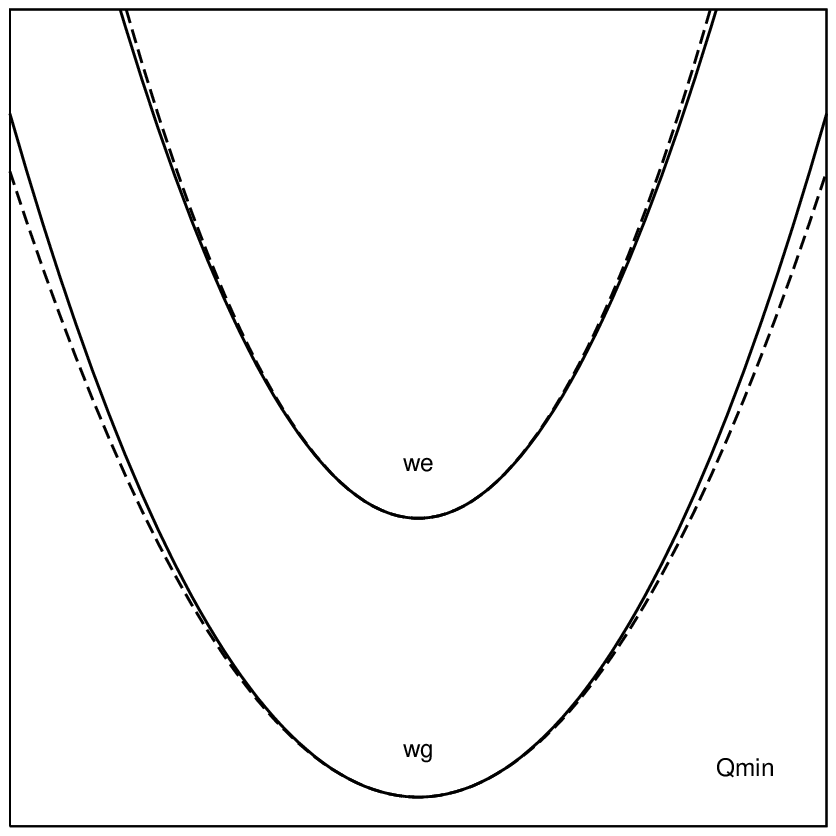} &\hspace*{0.5cm}&
        (c)\, \includegraphics[width=2.1cm]{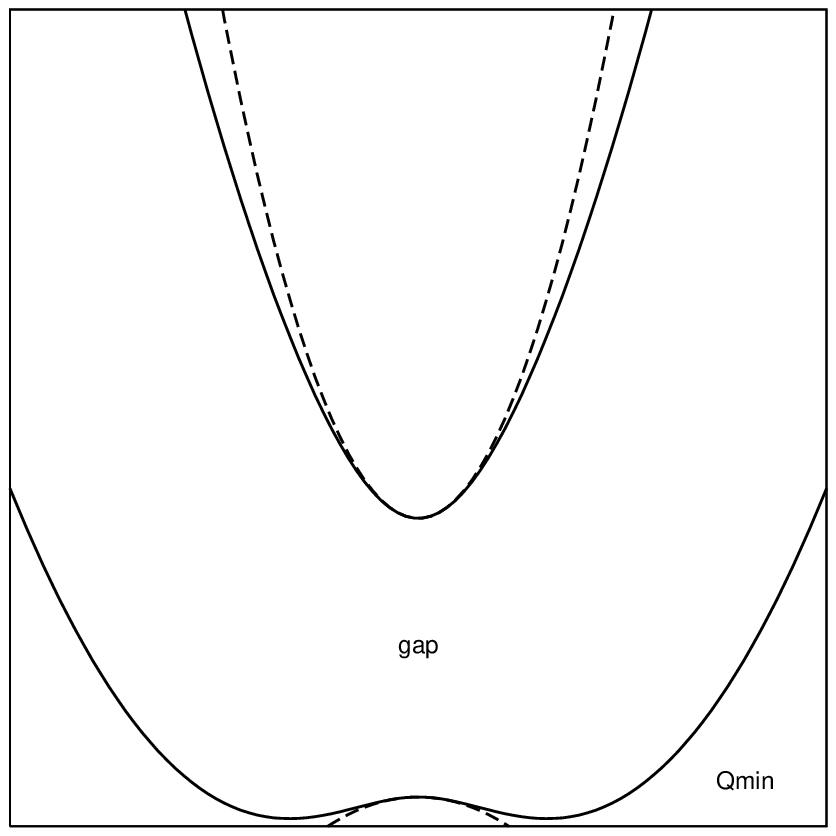} \\
      \end{tabular}\\
    \end{tabular}\\
    \end{tabular}
    \caption{
    (Color online)
    (a) Sketch: dimer molecule trapped in a nanogap
    between two voltage biased electrodes, $V_b = \mu_L-\mu_R$,
    and capacitively coupled to a back-gate (not shown) at voltage $V_g$, which shifts the effective molecular energy levels.
    The monomers (blue) can vibrate along the local totally symmetric breathing mode
    and are connected by a mechanically stiff bridge (not shown).
    (b-c) Adiabatic potentials $W_{g}$ and $W_{e}$ for the symmetry-breaking molecular distortion $Q_{-}$ (full lines)
    and harmonic expansions around $Q_{-}=0$ (dashed lines)
    for (b) weak ($\lambda=0.7$, $t=2.15\omega$) and (c) intermediate strength of the pJT effect ($\lambda=1.93$, $t=2.15\omega$).
    }
    \label{fig:setup}
  \end{figure}
It thereby significantly distorts the occupied monomer along coordinate $Q_i$ due to a change of the bond-lengths (c.f. Fig.~\ref{fig:setup}(a)).
The resulting shift of the potential minimum  is $\sqrt{2}\lambda$, expressed in units of the zero-point motion energy of the vibration of the undistorted monomers.
Thus $\lambda$ is the dimensionless electron-vibration coupling.
This model applies, for example, to a mixed-valence molecule,
where the monomers are metal-ions (zero spin) with a surrounding shell of ligand atoms.
%
% SUMMARY OF THE MODEL IN WORDS
Because of its coupling to the presence of the hopping excess electron, the
local distortion of the monomer is dragged along by the electron and becomes similarly delocalized over the dimer.
This results in coherent electro-mechanical motion and the breakdown of the Born-Oppenheimer separation.
The central quantity controlling the character of the molecular states is
the delocalization energy $t$ relative to the coupling $\lambda \omega$ to the localized distortion.
We note that in ref.~\cite{Kaat05}, the opposite case of stiff monomers and a distorted bridge was considered,
requiring only a single-mode to be considered, whereas here we account for two vibrational modes and their interplay.
Also, below we consider transport up to high bias $V_b\sim 4t$, in contrast to~\cite{Donarini06}.
When the size of the nanogap is varied,
which is possible in a mechanically controllable break junction,
the intra-molecular hopping, $t$, will change.
This allows for an in situ change of the character of the molecular state.
In principle, changes in $\omega$ and $\lambda$ may also occur,
but they do not alter the results qualitatively
and can safely be neglected.
\\
We assume charging effects (Coulomb blockade) to be strong enough that only two molecular 
charge states participate in transport processes, which we label by the number of excess electrons on the molecule $N=0,1$.
The Hamiltonian $H^N$ for the molecule in charge state $N$,
written in the molecular vibrational coordinates $Q_\pm=(Q_{\mathrm{2}} \pm Q_{\mathrm{1}})/\sqrt{2}$, then reads
\begin{eqnarray}
  H^0&=&\sum_{j=\pm} \half \omega \left( P_j^2+Q_j^2\right)
		 \label{eq:H_vib} \\
  H^1&=& H^0
                   - \lambda\omega Q_+
                   + \lambda\omega Q_- (\hat{n}_{1}-\hat{n}_{2}) \nonumber \\
	        & &+ t \sum_{\sigma} ( d_{1\sigma}^\dagger d_{2\sigma} +h.c.)
		 \label{eq:H_vib_pjt}
\end{eqnarray}
where $d_{i\sigma}^\dagger$ creates an electron in state $\vert i \sigma \rangle$
and $\hat{n}_i=\sum_{\sigma}d_{i\sigma}^\dagger d_{i\sigma}$ is the occupation operator.
The symmetric coordinate, $Q_{+}$, corresponds to the monomer vibrating in phase i.e.
 the \emph{molecular} breathing mode where the molecule as a whole changes its size.
It couples to the total excess charge, $\propto N$, of the molecule,
resulting in a simple shift of the potential surface along $Q_{+}$ by an amount $\lambda$ (linear term in eq.~(\ref{eq:H_vib_pjt})).
The Franck-Condon (FC) transport effects resulting from this type of coupling have been calculated by several groups~\cite{Braig03a,Mitra04b,Koch04b,Wegewijs05}
and found experimentally~\cite{Park00,Pasupathy04,Osorio07a}.
 % What is new
In contrast, the anti-symmetric mode, $Q_{-}$, corresponds to the monomers vibrating with opposite phase.
If an excess electron is present, this molecular shape distortion couples to the internal charge imbalance $\hat{n}_{1}-\hat{n}_{2}$.
Due to the intra-molecular tunneling, $t$, the Hamiltonian~(\ref{eq:H_vib_pjt}) mixes electronic and vibrational states of the mode $Q_{-}$
prohibiting a factorization of the molecular wave function into a $Q_{-}$-vibrational and an electronic part.
We thus need \emph{vibronic} states,
$\vert m_{-{}}, \sigma \rangle = \vert \chi_{m_{-}}^1 \rangle \vert 1 \sigma \rangle+\vert \chi_{m_{-}}^2 \rangle \vert 2 \sigma \rangle$,
to describe the excess electron and the pJT-active mode.
Here $m_{-}$ denotes the vibronic quantum number for the joint electron-vibration ($Q_{-}$) system:
distinguishing these systems is fundamentally impossible due to the quantum coherence.
Finding the vibrational coefficients $\vert \chi_{m_{-}}^i \rangle$ in the molecular {vibronic} eigenstates
requires diagonalization of the Hamiltonian~(\ref{eq:H_vib_pjt}), which has to be done numerically.
Despite the breakdown of the BO-approximation,
it is instructive, e.g. for interpreting the energy spectrum,
to consider the adiabatic potentials for the $Q_{-}$ vibrations,
obtained  by neglecting the nuclear kinetic energy operator ($P_{-} \rightarrow 0$) in eq.~(\ref{eq:H_vib_pjt})
and to find the electronic eigenstates as function of $Q_{-}$, while neglecting $Q_{+}$.
The resulting electronic energies are the ground ($g$) and excited ($e$) adiabatic $Q_{-}$-potential for the vibrations,
$W_{g,e}(Q_{-})=\half\omega Q_{-}^2\mp \sqrt{(\lambda\omega Q_{-})^2+t^2}$,
which are sketched in fig.~\ref{fig:setup}(b-c).
However, in the section \textit{Results} we will discuss
when the breakdown of the BO-approximation renders these potentials meaningless.

\section{Transport}
The transport setup, sketched in fig.~\ref{fig:setup} (a), is described by the Hamiltonian
$H = H_{\mathrm{M}} + H_{\mathrm{res}}+H_{\mathrm{T}}$.
Here the molecular part $H_{\mathrm{M}}$ is specified by
$H^N-\alpha V_g N$
for $N=0,1$ excess electrons
where the gate voltage $V_g$ linearly shifts the addition energies (proportional to the gate coupling $\alpha$).
The electrodes $r=L,R$ are described by
$
  H_{\mathrm{res}} = \sum_{r,k,\sigma} \left( \epsilon_{k}-\mu_r \right) c_{rk\sigma}^\dagger c_{rk\sigma}
$
and are kept at temperature $T$ and electrochemical potentials $\mu_r=\mu\pm V_b/2$
i.e. $V_b$ is the bias voltage.
They are coupled to the molecule by
$
  H_{\mathrm{T}} =  \sum_{r, i, k,\sigma} T_{r}^i d_{i\sigma}^\dagger c_{rk\sigma} + h.c.,
$
where $c_{rk\sigma}^\dagger$ creates an electron of spin $\sigma$ in state $k$ in electrode $r$
and $T_r^i$ is the amplitude for tunneling between the electrode $r$  and monomer $i=1,2$.
We consider a linear  arrangement as sketched in fig.~\ref{fig:setup}(a) by assuming $T_L^2=T_R^1=0$
and, for simplicity, symmetric coupling $T_L^1=T_R^2= \sqrt{\Gamma / (2\pi \rho)}$.
Here $\Gamma$ denotes the tunneling rate and $\rho$ is the electrode density of states.
In many experiments~\cite{Park00,Pasupathy04,Heersche06,Jo06,Osorio07a,Osorio07b} the transport is dominated by single-electron tunneling. We therefore focus on the corresponding range of applied voltages and temperature.
We calculate the non-equilibrium stationary state occupations of the molecular states
and the transport current both up to second order in $H_{\mathrm{T}}$ using a standard kinetic (master) equation, see e.g.~\cite{Wegewijs05}.
We have checked by explicit calculation that the off-diagonal elements of the density matrix are not crucial.
Rather it is more important to take a large number of states into account (order of $1000$ states).
\begin{figure}[htbp]
  \centering
  \psfrag{Vb}[r][r]{$V_b[\omega]$}
  \psfrag{Vg}[c][c]{$\alpha V_g[\omega]$}
  \psfrag{(i)}[l][l][0.7][0.0]{(i)}
  \psfrag{(ii)}[l][l][0.7][0.0]{(ii)}
  \psfrag{(iii)}[b][b][0.7][0.0]{(iii)}
  \psfrag{gap}[t][t][0.7][0.0]{gap}
  \psfrag{cb}[l][l]{$\frac{dI}{dV_b}[\frac{\Gamma}{\omega}]$}
  \begin{tabular}{c}
    \includegraphics[width=7cm,height=6cm]{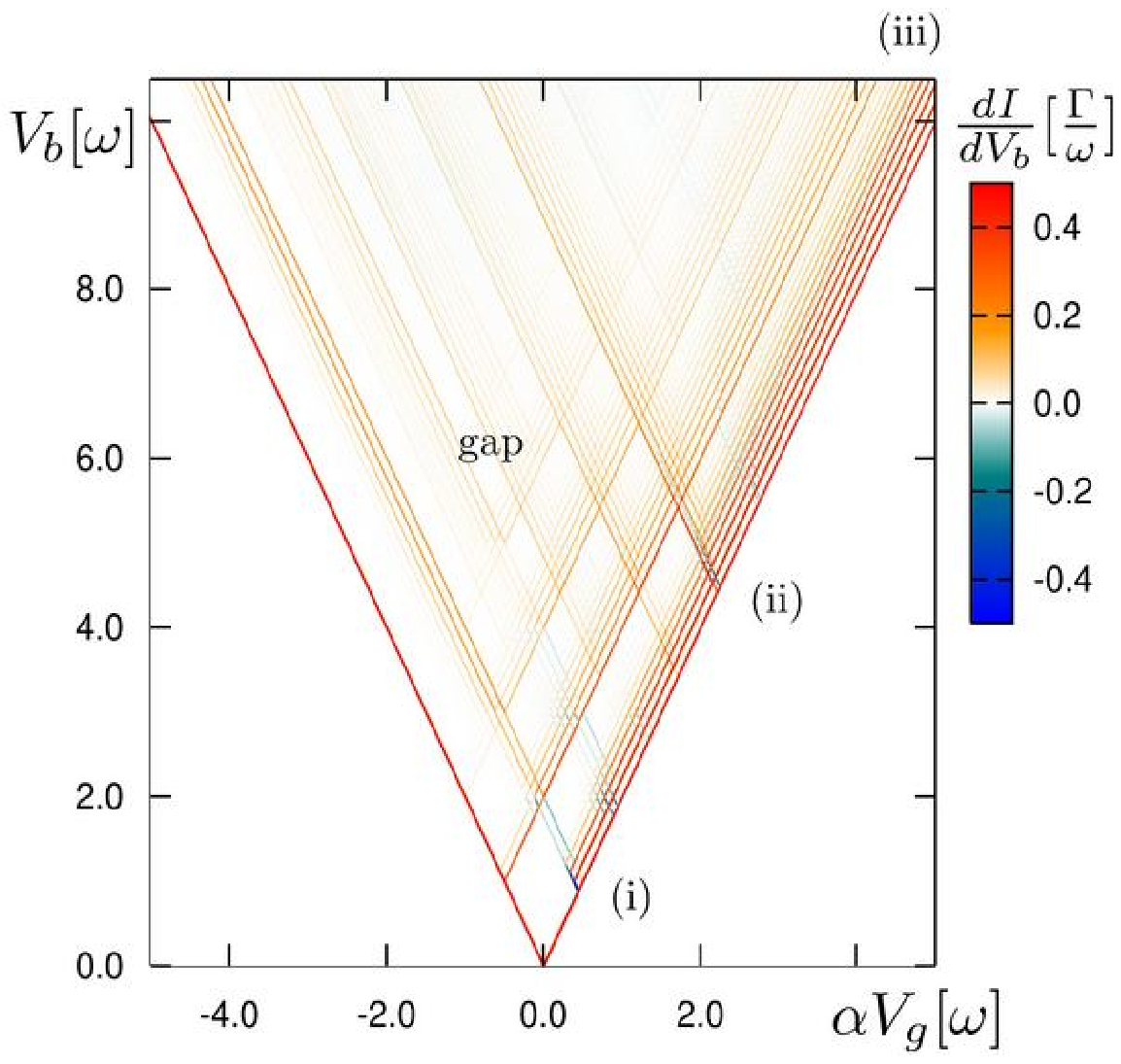}\\(a)\vspace{0.5cm}\\
    \includegraphics[width=7cm,height=6cm]{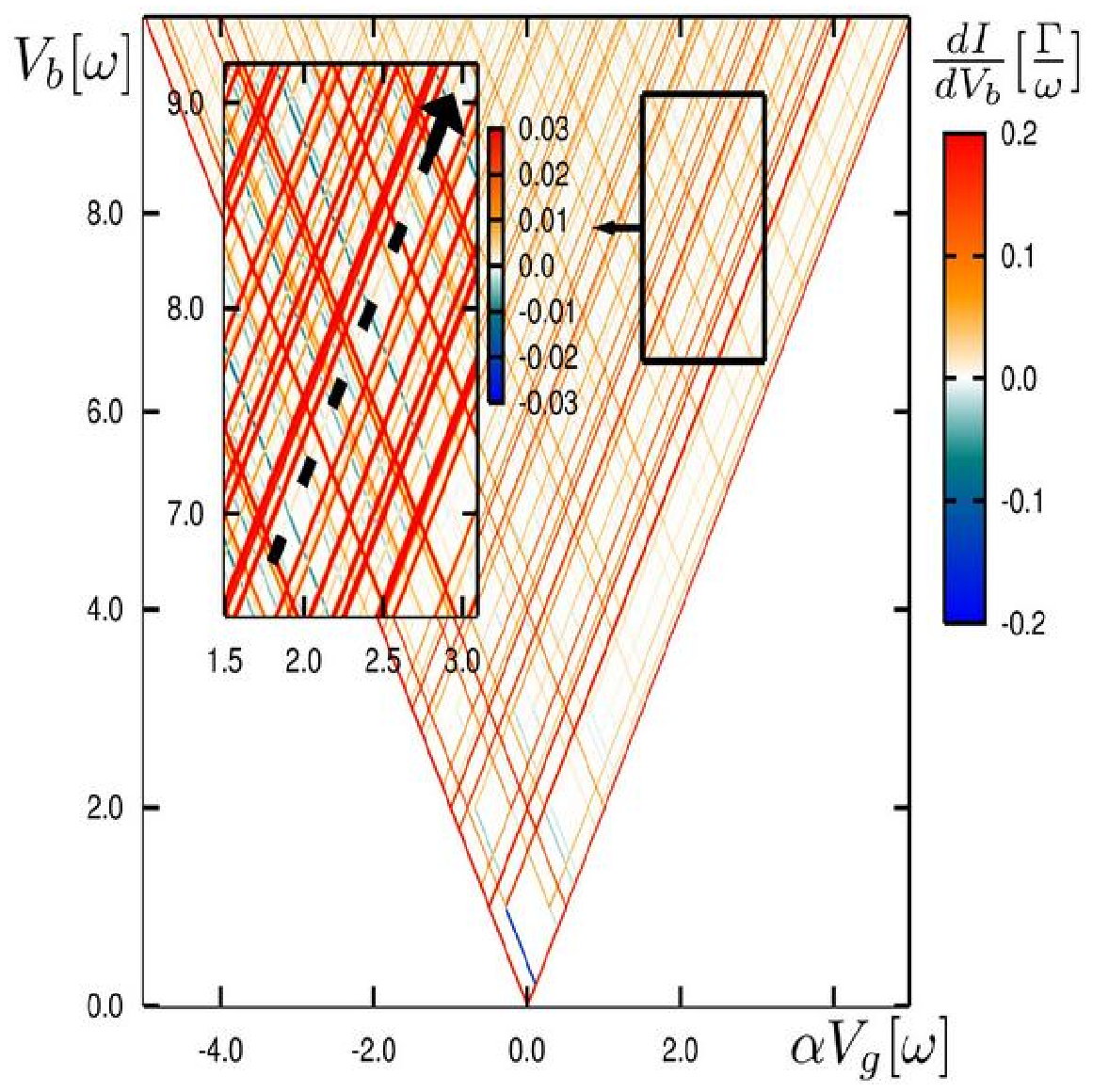}\\(b)
  \end{tabular}
  \caption{
    (Color online)
    $dI/dV_b$ ($\Gamma=2.5 \cdot 10^{-5}\omega$, $T=4\cdot 10^{-3}\omega$) for
    (a) weak pJT mixing ($\lambda=0.7$, $t=2.15\omega$)
    (b) moderate  pJT mixing ($\lambda=1.93$, $t=2.15\omega$).
    Inset: high contrast, dashed black line marking the $V_b-V_g$ trace taken in fig.~\ref{fig:strong}(a).
    For convenience the gate voltage is defined such that  $V_g=0$ corresponds to the charge degeneracy point.
    Due to symmetric biasing, energy scales appear at twice the separation on the voltage axis.
  }
  \label{fig:weak}
\end{figure}
\begin{figure}[htbp]
  \psfrag{E}[r][c]{$E^{1}_{-} [\omega]$}
  \psfrag{t}[c][c]{$t [\omega]$}
  \psfrag{Vb}[r][c]{$V_b[\omega]$}
  \psfrag{cb}[l][l]{$\frac{dI}{dV_b}[\frac{\Gamma}{\omega}]$}
  \psfrag{0}[r][c][0.7][0.0]{$0$}
  \psfrag{2}[r][c][0.7][0.0]{$2$}
  \psfrag{4}[r][c][0.7][0.0]{$4$}
  \psfrag{6}[r][c][0.7][0.0]{$6$}
  \psfrag{8}[r][c][0.7][0.0]{$8$}
  \psfrag{10}[r][c][0.7][0.0]{$10$}
  \psfrag{0.5}[c][c][0.7][0.0]{$0.5$}
  \psfrag{1.0}[c][c][0.7][0.0]{$1.0$}
  \psfrag{1.5}[c][c][0.7][0.0]{$1.5$}
  \psfrag{2.0}[c][c][0.7][0.0]{$2.0$}
  \psfrag{2.5}[c][c][0.7][0.0]{$2.5$}
  \psfrag{3.0}[c][c][0.7][0.0]{$3.0$}
  \psfrag{3.5}[c][c][0.7][0.0]{$3.5$}
  \centering
  \begin{tabular}{c}
    \includegraphics[width=7.5cm, height=6cm]{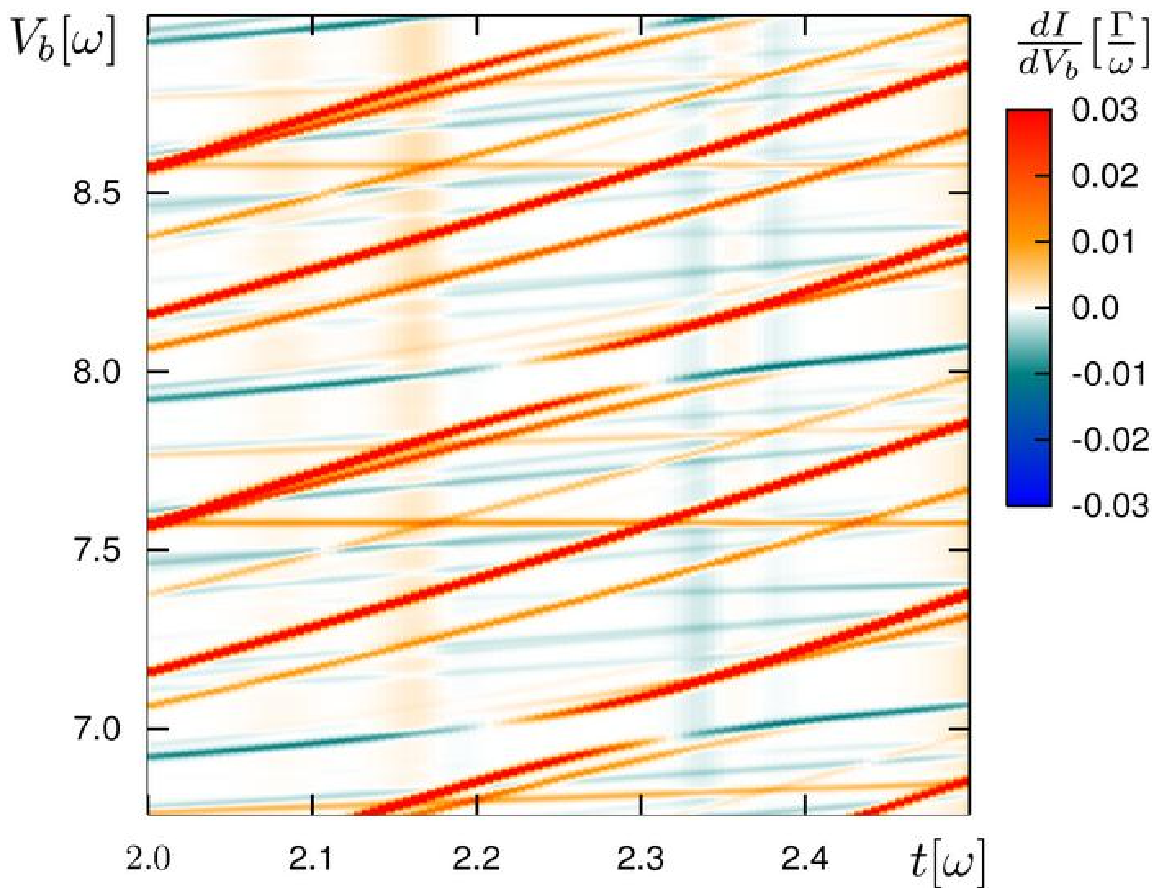}\\(a)\vspace{0.4cm}\\
    \includegraphics[width=5.5cm,height=5.5cm]{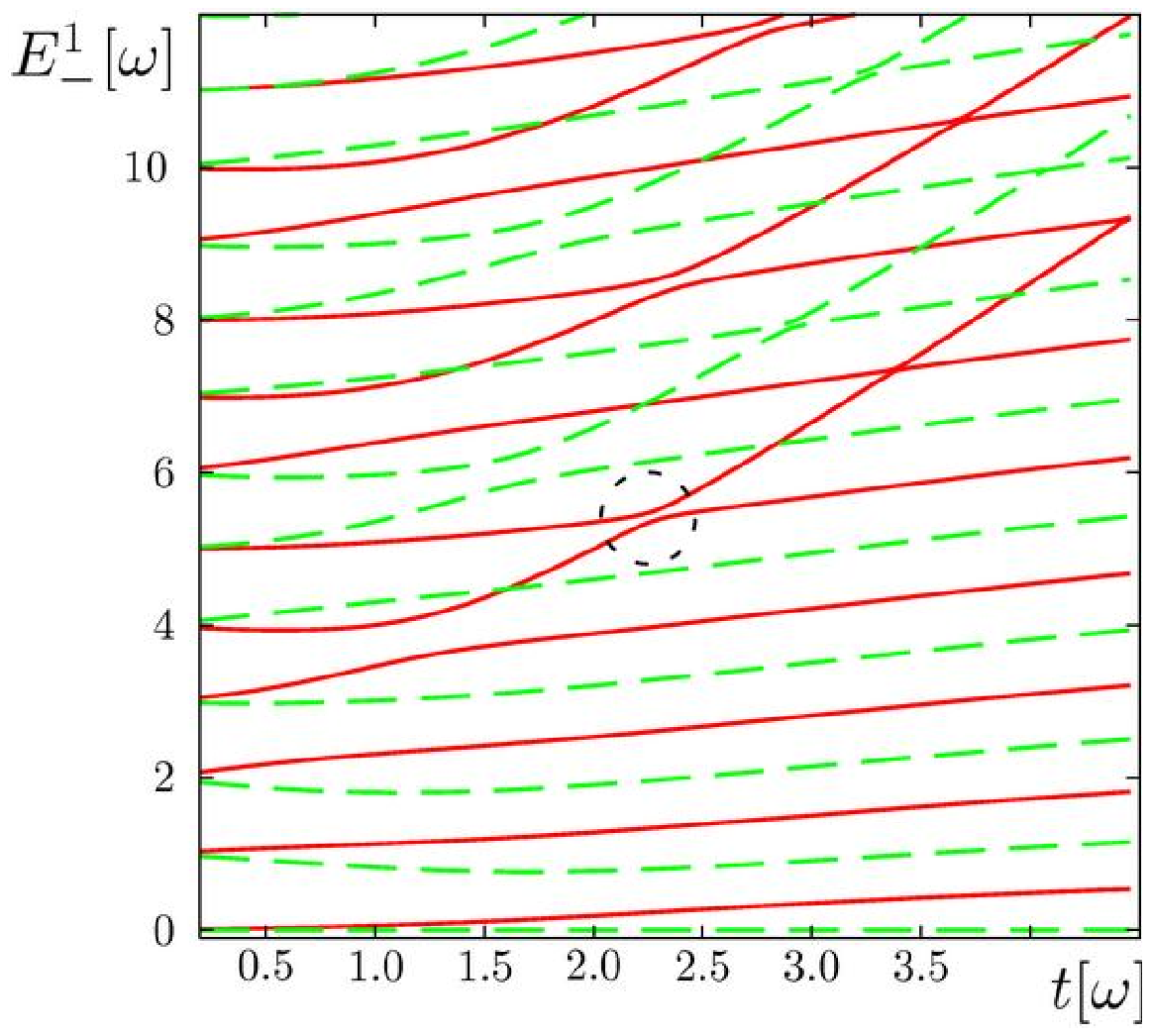}\\(b)\vspace{0.4cm}\\
    \includegraphics[height=3cm,width=7cm]{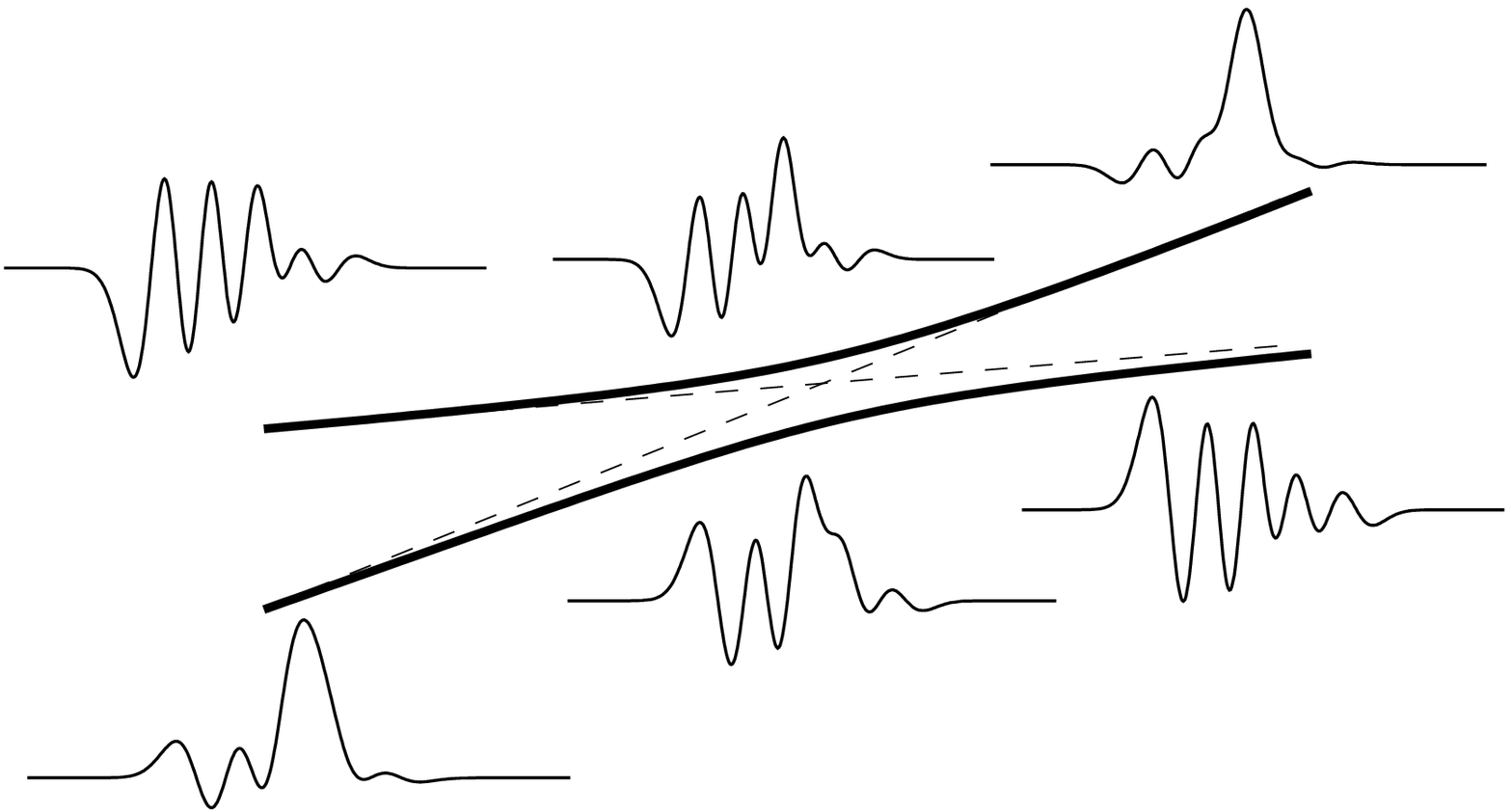}\\(c)
  \end{tabular}
  \caption{
    (Color online)
    Signature of the pJT effect:
    anti-crossings as the intramolecular delocalization $t$ is varied due to a mechanical change of the nanogap size.
    (a) Evolution with $t$ of the $dI/dV_b$ trace along the line in the $V_b,V_g$ plane marked in the inset of fig.~\ref{fig:weak}(b).
    (b) Evolution with $t$ of energies of the vibronic states for $N=1$ ($\lambda=1.93$),
        with the harmonic $Q_{+}$-vibration energies subtracted.
        The green (dashed) / red (solid) color indicates positive / negative parity.
        The anti-crossing in the transport in (a) corresponds to the marked anti-crossing around $t \approx 2.3\omega$ and $E^1_{-}\approx 5\omega$.
    (c) Evolution with $t$ of the vibrational parts $\chi^1_{m_{-}}(Q_{-})$
    for $m_{-}=9$ and $10$, respectively.
        Due to the symmetry of the dimer, the vibronic state has  definite molecular parity $\pi=\pm$,
        reflected by the property $\chi_{m_{-}}^1(Q_{-})=\pi  \chi_{m_{-}}^2(-Q_{-})$.
  }
  \label{fig:strong}
\end{figure}
\\
\section{Results}
In order to appreciate the breakdown of the BO separation, we first discuss a case where it has approximate validity
i.e. the effect of the distortion of the monomers is sufficiently weak.
In fig.~\ref{fig:weak}(a) we show the differential conductance as function of the applied voltages.
% First mention the well-known features
Many excitations appear, involving the $Q_{+}$ and/or the $Q_{-}$ mode, which are separated in bias voltages by multiples of $2\omega$ (due to symmetric biasing).
The first of these excitations starts out at the marker (i) in fig.~\ref{fig:weak}(a).
% New stuff 
However, in contrast to usual FC transport spectra~\cite{Mitra04b,Koch04b,Wegewijs05},
the $Q_{-}$ excitations are weakened within the noticeable gap of $4t$
(due to exponentially suppressed overlap integrals of classically forbidden transitions),
and enhanced conductance peaks delimit the upper boundary of the gap (starting out from marker (ii)).
One can thus directly estimate the strength of the delocalization of the excess electron.
Furthermore, the excitations spaced in $V_b$ roughly by $2\omega$ also have a detailed substructure of a dense series of conductance peaks,
for instance, along the right edge of the transport region, terminating at the marker (iii).
The latter correspond to tunnel processes between $Q_{-}$ excitations of the $N=0$ and $N=1$ ground potential
with the same vibrational quantum number $m_{-}=0,1,2,\ldots$.
For $\lambda^2\omega \ll t$, the potential $W_{g}(Q_{-})$ is approximately harmonic,
but with reduced frequency $\omega_g/\omega \approx 1-\omega\lambda^2/t$ due to the pJT interaction,
see fig.~\ref{fig:setup}(b).
Therefore the resonances corresponding to different $m_{-}$ occur  at slightly different positions~\cite{Wegewijs05}.
The equidistant energy spacings correspond to $\omega -\omega_g\approx (\lambda\omega)^2/t \ll \omega$.
Similarly, above the gap near marker (ii), a dense series of conductance peaks with negative $V_g$-dependence indicates that
 the upper adiabatic potential has a higher frequency $\omega_{e}/\omega =1+\omega\lambda^2/t$.
Using the gap and features at (i)-(iii) in  fig.~\ref{fig:weak}(a)
one can thus estimate $\lambda$, $t$ and $\omega$ from the transport data.
For larger values of $\lambda$, the adiabatic potentials additionally become anharmonic due to the pJT interaction
resulting in markedly non-equidistant spacing of the dense series of conductance peaks.
We note that, interestingly, in mixed-valence molecules with magnetic ions this vibrational an-harmonicity
correlates with the relative orientations of the ionic spins~\cite{Reckermann09a}.
\\
The most dramatic effect of the pJT interaction is the breakdown of the BO separation for stronger coupling $\lambda$.
This occurs when \emph{excited states} of the two adiabatic potentials $W_e$ and $W_g$ come close in energy.
These anti-crossings can be observed in the transport
at high bias voltages, $V_b \gtrsim 4 t$ (c.f. fig.~\ref{fig:setup}(c)), where the states of the excited adiabatic potential become accessible as well.
At a first glance, the transport spectrum in this case, shown in fig.~\ref{fig:weak}(b), seems inextricably complex.
However, clear signatures of the pJT effect are revealed when the nano-gap size is varied and the intramolecular hopping $t$ changes
 while the mechanical properties of the monomers $\lambda,\omega$ remain fixed.
In fig.~\ref{fig:strong}(a) we show the evolution of the $dI/dV_b$ trace taken along the dashed black line in the inset of fig.~\ref{fig:weak}(b),
as $t$ is varied.
Experimentally, such data can be collected with techniques described in~\cite{Champagne05,Parks07}. 
Among the most pronounced $dI/dV_b$ resonances, those with a weak $t$-dependence one would assign to highly excited vibrations in the lower adiabatic electronic state, $W_g$,
whereas those with a strong linear $t$-dependence would correspond to the lowest excitations in the upper adiabatic electronic state, $W_e$.
The main difference in $t$-dependence stems from the gap, $~ 2t$, separating the adiabatic potentials (c.f.~\ref{fig:setup}(c)).
This distinction is completely lost at the anti-crossings visible in fig.~\ref{fig:strong}(a).
The conductance anti-crossing at $t \approx 2.3\omega$ maps out the
corresponding anti-crossing in the evolution of the vibronic energy spectrum with $t$ which is shown in fig.~\ref{fig:strong}(b).
Strikingly, the \emph{sign} of the conductance of the two anti-crossing transport resonances is different.
This directly relates to the large difference in the kinetic energy of the nuclear motion of the two anti-crossing adiabatic electronic states.
In fig.~\ref{fig:strong}(c) we show the real space representation of vibrational components of the
vibronic wave functions of the involved states.
The weakly $t$-dependent excitation has a rapidly varying wave function
leading to a small overlap with the vibrational groundstate of the uncharged molecule.
The occupation of this state on average reduces the contributions to the current of other states, therefore leading to negative differential conductance~\cite{Wegewijs05}.
In contrast, the strongly $t$-dependent excitation is more similar in shape to the vibrational groundstate leading to a much larger overlap integral and therefore positive differential conductance.
At the anti-crossing the strong pJT mixing causes the components of \emph{both} vibronic wave functions to rapidly vary.
As a result the conductance peaks disappear in a narrow range of $t$ values in the anti-crossing region.
Note that all other resonances, which follow from the BO approximation and the FC-principle, smoothly depend on $t$,
making the pJT effect clearly stand out.
Strikingly, the transport anti-crossings seen in fig.~\ref{fig:strong}(a) are \emph{replicas} of one and the {same} anti-crossing marked in fig.~\ref{fig:strong}(b),
due to the simultaneous excitation of the $Q_{+}$ mode.
This can be seen from both the voltage distance of the anti-crossings and the identical $t$-dependence.
Thus, interestingly, the \emph{pJT-inactive mode} proliferates the violation of the adiabatic BO separation in the transport.
The many other anti-crossings in fig.~\ref{fig:strong}(b) result in pJT resonances at different voltages (not shown),
and comparison with these calculated levels allows one to estimate the parameters.
More generally, the effects exemplified above may be expected whenever the pJT mixing is important,
 that is, for minimal separation of the adiabatic potentials on the order of the vibrational quanta, $t \sim \omega$,
and moderate to strong coupling to the distortion, $\lambda \gtrsim 1$.

\section{Conclusion}
In this letter we have shown for a minimal model exhibiting the dynamical pJT effect,
that the breakdown of the Born-Oppenheimer separation of the electronic
and vibrational motion in a molecular transistor leads to novel
transport resonances which can be distinguished from standard Franck-Condon effects.
The combination of electrostatic gating and mechanical control is crucial to unravel such complex molecular transport processes
and demonstrate electro-mechanical quantum entanglement in molecule-based NEMS.
Interesting candidate devices are mixed-valence molecules with a moderate degree of intramolecular delocalization of the excess electron,
so-called Robin-Day Class II systems.
Their electron transport properties may shed new light on the fundamental issue in physical chemistry of their classification by
intramolecular charge transfer.

\acknowledgments
We acknowledge K. Flensberg  for discussions and the financial support from
DFG SPP-1243,
the NanoSci-ERA,
the Helmholtz Foundation,
the EU-RTN Spintronics,
and the FZ-J\"ulich (IFMIT).

\bibliographystyle{eplbib}
\bibliography{cite}

\end{document}